\RequirePackage{fix-cm}
\documentclass[smallextended]{svjour3}       % onecolumn (second format)
\smartqed  % flush rightStorage  qed marks, e.g. at end of proof

\usepackage[margin=1in]{geometry}
\usepackage{amsmath,amssymb,graphicx,booktabs,url}
\usepackage[hidelinks]{hyperref}
\usepackage{natbib}
\usepackage{graphicx}
\usepackage{mathptmx}
\usepackage{latexsym}
\newcommand{\Rpde}{\mathcal{R}_{\mathrm{PDE}}}
\newcommand{\Rstr}{\mathcal{R}_{\mathrm{str}}}
\newcommand{\Rbc}{\mathcal{R}_{\mathrm{bc}}}
\newcommand{\Rsol}{\mathcal{R}_{\mathrm{sol}}}     % use Times fonts if available on your TeX system

\journalname{Machine Learning: Science and Technology}

\begin{document}
	
	\title{Inverted model selection in physics-informed neural networks}
	\subtitle{when a lower residual selects a worse solution}
	
	%\titlerunning{Short form of title}        % if too long for running head
	
	\author{Rabiu Musah         %\and
	%	Second Author %etc.
	}
	
	%\authorrunning{M. Rabiu} % if too long for running head
	
	\institute{M. rabiu \at
		Department of Applied Physics, Faculty of physical Sciences, University for Development Studies,\\ 
		Tamale, Ghana \\
		Tel.: +233-54-5843562\\
		%Fax: +123-45-678910\\
		\email{mrabiu@uds.edu.gh}           %  \\
		%             \emph{Present address:} of F. Author  %  if needed
		%\and
		%S. Author \at
		%second address
	}
	
	\date{Received: \today() / Accepted: date}

	\maketitle
	\begin{abstract}
		Physics-informed neural networks (PINNs) are commonly compared, tuned and reported through a single aggregate residual. That practice rests on an assumption which, to my knowledge, has not been tested directly: that a smaller residual indicates a better solution. Across three constrained PDE systems I find the assumption can fail, and fail systematically. In matched pairs of solvers differing only in whether a defining structural identity is hard-wired or imposed as a penalty, the penalised variant frequently attains the lower equation residual while violating that identity by several orders of magnitude; the variant satisfying the identity exactly is then ranked worse. Over 64 matched pairs spanning two systems, four network variants and eight seeds, this inversion occurs in 83\% of cases (95\% Wilson CI 72--90\%), though not universally: rates run from 72\% to 94\% across the two systems. Repeating the comparison across six  architecture families---MLP, cPINN, XPINN, hp-VPINN, and physics-informed DeepONet and FNO---inverts in 46 of 48 pairs, indicating that the variability is problem-dependent rather than specific to the approximator. A third, larger vorticity--streamfunction problem shows the same ordering, with the residual-optimal solver violating its structural identity by more than six orders above tolerance, though there the residual margin between the two is only 9.37\%. Because a scalar loss cannot expose this, I pair the diagnosis with a lexicographic admissibility gate---structural identity, boundary trace, and a solvability integral that must vanish independently of the equation residual---and require all three before residuals are permitted to 	compete. The gate catches three artefact classes but demonstrably not a fourth: a prescribed-structure prior yields fields that pass every single-run check, and only deleting the source term reveals that $97\%$ of the reported structure survives removal of the physics. Reference data and figures accompany the paper.
		\keywords{physics-informed neural networks, verification, benchmark, constrained PDE, solution admissibility, scientific machine learning}
		\subclass{65M75 \and 68T07 \and 65M60}
	\end{abstract}

	\section{Introduction}
	
	Reported success in the PINN literature is usually mediated by a single number: a training loss, or a residual norm evaluated after convergence. By the conventions of computational physics this is an unusually thin standard. A finite-element or finite-volume code is normally accepted only after it has been shown to respect the constraints of the system it claims to solve---incompressibility, boundary data, conservation laws---to a 	stated tolerance, on data withheld from fitting.
	
	Whether that difference matters is an empirical question, and it is the question this paper addresses. Most PDE systems of physical interest are constrained, and when a constraint enters as a penalty rather than by 	construction, the composite objective admits minimisers that trade constraint violation against equation residual. There is no guarantee that the trade settles somewhere harmless. A scalar loss, in any case, cannot 	report where it settled.
	
	The experiments reported below indicate that the trade often settles badly, and in a particular direction. Comparing solvers that differ only in how a defining identity is imposed, the penalised variant tends to reach a lower equation residual precisely because it is free to violate the identity; ranking by residual then prefers it. Following the relaxation argument of Section~\ref{sec:theory}, this is what one should expect when the penalty weight is small; I refer to it as an inversion of model selection. On the systems examined here it is common rather than exceptional, occurring in roughly six of every seven matched pairs---but it is not universal, and one of the three systems inverts only about seven times in ten, which I take as evidence that the effect depends on the problem in ways not yet understood.
	
	Section~\ref{sec:related} places the work against existing accounts of PINN failure. Sections~\ref{sec:bench} and \ref{sec:gate} introduce the benchmark and the gate built from its checks. The artefact classes appear in Section~\ref{sec:results} and the generality experiments in Section~\ref{sec:general}. Section~\ref{sec:theory} gives the relaxation argument. What I could not settle is collected in Section~\ref{sec:open}, and Section~\ref{sec:limits} closes with two limitations that appear inherent.
	
	A note on provenance: none of the four artefact classes was invented for this paper. Each was encountered while developing a PINN solver for a magnetohydrodynamic vortex problem, where they were initially mistaken for 	convergence difficulties. The benchmark was constructed afterwards, to isolate and reproduce them under controlled conditions.

	\section{Related work}\label{sec:related}
	
	The PINN formulation of \citet{raissi2019} embeds the residual of a governing equation in the training objective, with boundary and initial data entering as additional penalty terms. Difficulties with that formulation are by now well documented. \citet{krishnapriyan2021} show that PINNs fail to recover the correct solution for convection and reaction problems whose finite-difference treatment is unremarkable, and trace the failure to the optimisation landscape rather than to expressivity. \citet{wang2021gradient} identify a stiffness in the gradient flow when loss terms of differing scale are combined, and \citet{wang2022ntk} reach a similar conclusion through the neural tangent kernel, showing that the effective convergence rate of each term depends on its weight. My observations are consistent with that literature, but concern a different question: not whether the optimiser reaches a good solution, but whether the reported residual allows one to tell.
	
	Removing constraints from the objective altogether is an established remedy. \citet{lu2021hard} impose hard constraints for inverse design, and \citet{sukumar2022} construct distance functions that satisfy boundary data exactly. The structural formulation used here belongs to that family, and my contribution is to quantify what the penalised alternative costs in model selection rather than in accuracy alone.
	
	Architectural variants---cPINN \citep{jagtap2020cpinn}, XPINN \citep{jagtap2020xpinn}, hp-VPINN \citep{kharazmi2021}---and operator-learning methods such as DeepONet \citep{lu2021deeponet} and Fourier neural operators \citep{li2021fno} address other limitations of the basic formulation. None is examined here, for reasons given in Section~\ref{sec:general}.
	
	The practice advocated here is standard elsewhere in computational physics. Code verification through grid convergence and the method of manufactured solutions \citep{roache2002}, and the broader verification and validation framework of \citet{oberkampf2010}, both rest on the principle that a solver must be shown to satisfy the constraints it claims to enforce before its accuracy is discussed. Very little of that apparatus has migrated into the PINN literature.
	
	\section{Benchmark problem}\label{sec:bench}
	
	On the unit disk $\Omega=\{r<1\}$ I solve the steady forced vorticity transport problem
	\begin{align}
		\nu\nabla^{2}\omega-\mathbf{u}\cdot\nabla\omega+S(r)&=0, \label{eq:pde}\\
		\omega+\nabla^{2}\psi&=0, \label{eq:str}\\
		\mathbf{u}&=(\partial_y\psi,\,-\partial_x\psi), \label{eq:vel}\\
		\psi&=0 \quad\text{on } r=1, \label{eq:bc}
	\end{align}
	with $\nu=2\times10^{-2}$ and an annular source $S(r)=A\exp[-((r-r_0)/w)^2]$, $A=1$, $r_0=0.60$, $w=0.12$. This is referred to as problem S1.
	
	The problem is deliberately minimal, but it carries three checks that are logically independent of \eqref{eq:pde}:
	
	\paragraph{(i) Structural identity} Equation~\eqref{eq:str} couples the two fields. A solver may output $\psi$ and $\omega$ independently and enforce \eqref{eq:str} by penalty, in which case $\Rstr=\langle(\nabla^{2}\psi+\omega)^{2}\rangle$ measures a genuine violation; or it may define $\omega:=-\nabla^{2}\psi$, in which case $\Rstr\equiv0$ by construction.
	
	\paragraph{(ii) Boundary trace} $\Rbc=\langle\psi^{2}\rangle_{r=1}$, evaluated on a rim sample disjoint from the training set.
	
	\paragraph{(iii) Solvability identity} With $\mathbf{u}$ given by \eqref{eq:vel}, the advection term is a Jacobian, $\mathbf{u}\cdot\nabla\omega=\{\psi,\omega\}$, whose integral over $\Omega$ vanishes. Hence
	\begin{equation}
		\Rsol=\Big|\int_\Omega \mathbf{u}\cdot\nabla\omega\,\mathrm{d}A\Big|
		\label{eq:sol}
	\end{equation}
	must vanish for any admissible field, independently of whether \eqref{eq:pde} is satisfied. The check is inexpensive---one quadrature over the collocation sample, at negligible cost relative to a single training step---and requires no reference solution.
	
	\subsection{Reference solution}
	
\begin{figure}[t]
		\centering\includegraphics[width=\textwidth]{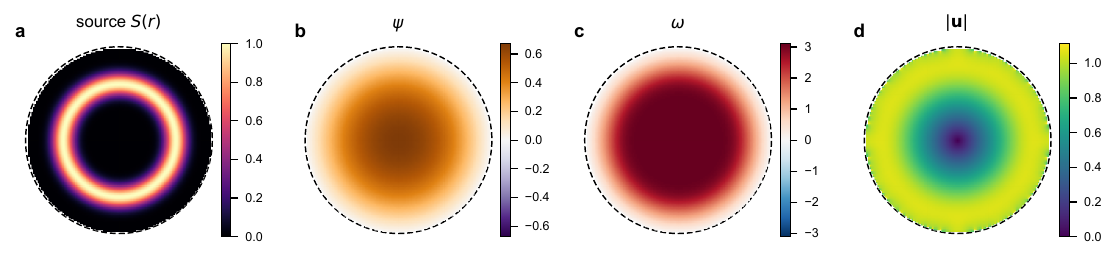}
		\caption{Benchmark problem and finite-difference reference: annular source, stream function, vorticity and speed.}\label{fig:ref}
	\end{figure}

	The reference is a second-order finite-difference discretisation on a masked disk with a direct sparse Poisson solve, relaxed to steady state. Because $\psi$ is obtained from $\omega$ by solving \eqref{eq:str}, the structural identity is imposed rather than penalised, and the Dirichlet trace \eqref{eq:bc} is imposed inside the solve. At $n=121$ the relaxed reference reaches an equation-residual root-mean-square of $3.62\times10^{-4}$ on its own grid. I write this $\|R\|_{\rm rms}$ to distinguish it from $\Rpde$, which throughout denotes the mean square of the same physical residual evaluated on held-out collocation points; the two differ by a square root and by sample, so the reference row of Table~\ref{tab:main} reads  $\Rpde=1.31\times10^{-7}\simeq\|R\|_{\rm rms}^{2}$. The reference satisfies $\Rstr=8.1\times10^{-25}$, $\Rbc=0$ and $\Rsol=1.8\times10^{-6}$ (Fig.~\ref{fig:ref}).
	
	\subsection{Verification of the reference solver}\label{sec:verify}
	
	Every admissibility judgement is made against the reference, so the reference itself requires code verification. Using the method of manufactured solutions with a stream function prescribed as,
	\begin{equation}
		\psi_e(x,y)=A\,(1-x^{2}-y^{2})^{4}\cos\tfrac{\pi x}{2}\cos\tfrac{\pi y}{2},
		\label{eq:mms}
	\end{equation}
	from which $\omega_e=-\nabla^{2}\psi_e$ and $\mathbf u_e$ follow, and the manufactured source $S_{\rm mms}=-\bigl(\nu\nabla^{2}\omega_e-\mathbf u_e\!\cdot\!\nabla\omega_e\bigr)$ is derived symbolically so that $\omega_e$ solves the continuous problem exactly. No derivative is taken by hand.
	
	The exponent in \eqref{eq:mms} is not decorative. The discretisation masks $\omega$ to zero outside the disk, so a manufactured field must satisfy both $\psi_e=0$ and $\nabla^{2}\psi_e=0$ on $r=1$; otherwise the boundary mismatch propagates through the elliptic solve and pollutes the interior. Writing $\psi_e=(1-r^{2})^{n}g$ gives $\nabla^{2}[(1-r^{2})^{n}]=-4n(1-r^{2})^{n-1}+4n(n-1)r^{2}(1-r^{2})^{n-2}$, which vanishes at $r=1$ for $n\ge3$ and any smooth $g$. Taking $n=4$ leaves $\omega_e=O((1-r^{2})^{2})$ near the rim, so the residual mismatch at the discrete mask $r=1-\Delta x$ is $O(\Delta x^{2})$ and does not degrade a second-order scheme. With $n=1$ the observed orders were $-0.06$: the errors grew under refinement, and restricting the norm to $r<0.7$ did not help, confirming that the contamination was global rather than a boundary layer.
	
	Errors are measured against $\omega_e$ in the $L^{2}$ and $L^{\infty}$ norms,
	\begin{equation}
		\|E_k\|_{L^{2}}=\sqrt{\frac{1}{N}\sum_{i=1}^{N}\bigl(\omega_i^{k}-\omega_{e,i}\bigr)^{2}},
		\qquad
		\|E_k\|_{L^{\infty}}=\max_i\bigl|\omega_i^{k}-\omega_{e,i}\bigr|,
		\label{eq:norms}
	\end{equation}
	with the empirical order of accuracy taken from successive refinements at constant ratio $r=\Delta x_k/\Delta x_{k+1}$,
	\begin{equation}
		p=\frac{\ln\bigl(\|E_k\|/\|E_{k+1}\|\bigr)}{\ln r}.
		\label{eq:order}
	\end{equation}
	Grid uncertainty is reported as the Grid Convergence Index
	\begin{equation}
		\mathrm{GCI}=\frac{F_s\,|\varepsilon|}{r^{p}-1},
		\qquad
		\varepsilon=\frac{f_k-f_{k+1}}{f_{k+1}},
		\label{eq:gci}
	\end{equation}
	with $F_s=1.25$ and $f_k$ the discrete $L^{2}$ norm of $\omega$ on grid $k$. The relative change is taken on a solution functional rather than on the error norm, since the GCI is intended to bound the uncertainty in a computed quantity without reference to an exact answer.
	
	\begin{table}[h]
		\centering\small
		\caption{Discretisation verification of the reference solver by manufactured solution. Grids are chosen for a constant refinement ratio $r=1.6$. The asymptotic order approaches the nominal second order of the central scheme from above.}
		\label{tab:grid}
		\begin{tabular}{lcccccc}
			\toprule
			$n$ & $\Delta x$ & $\|E\|_{L^{2}}$ & $p_{L^{2}}$ & $\|E\|_{L^{\infty}}$ &
			$p_{L^{\infty}}$ & GCI (\%)\\
			\midrule
			51  & $4.000\times10^{-2}$ & $2.743\times10^{-2}$ & ---  & $9.126\times10^{-2}$ & ---  & ---\\
			81  & $2.500\times10^{-2}$ & $9.290\times10^{-3}$ & 2.30 & $3.334\times10^{-2}$ & 2.14 & 1.414\\
			129 & $1.563\times10^{-2}$ & $3.654\times10^{-3}$ & 1.99 & $1.307\times10^{-2}$ & 1.99 & 0.893\\
			205 & $9.804\times10^{-3}$ & $1.448\times10^{-3}$ & 1.99 & $5.162\times10^{-3}$ & 1.99 & 0.527\\
			\bottomrule
		\end{tabular}
	\end{table}
	
	Both norms converge at 1.99 on the finest pair, against a nominal order of 	two, and the GCI falls from 1.41\% to 0.89\%. Two qualifications belong with 	the table. The verification is run diffusion-dominated ($\nu=1$, $A=1$), where the steady problem is solved by Picard iteration in thirteen sweeps; 	at the benchmark value $\nu=2\times10^{-2}$ the global P\'eclet number is 	large enough that the lagged-velocity iteration does not contract, and an explicit pseudo-time relaxation is limited by the advective step. The order of a scheme does not depend on $\nu$, but this study verifies the discretisation rather than the benchmark's operating regime. The asymptotic-range ratio, $\mathrm{GCI}_{21}/(r^{p}\mathrm{GCI}_{32})=0.62$, also indicates the coarsest grid is not yet fully asymptotic, consistent	with the observed order falling from 2.30 to 1.99 as the mesh refines.
	
	\section{The admissibility gate}\label{sec:gate}
	
	Every candidate is evaluated on a held-out sample of $6000$ interior points and $1024$ rim points, never the training collocation set, and apply	the following criterion.
	
	\begin{quote}
		A candidate is deemed \emph{admissible} if $\Rstr\le\varepsilon_{\rm str}$, $\Rbc\le\varepsilon_{\rm bc}$,
		$\Rsol\le\varepsilon_{\rm sol}$, and the enstrophy exceeds a non-triviality floor $Z_{\min}$. Only admissible candidates are ranked, and they are ranked by $\Rpde$.
	\end{quote}
	
	with $\varepsilon_{\rm str}=\varepsilon_{\rm bc}=10^{-6}$,
	$\varepsilon_{\rm sol}=5\times10^{-2}$, $Z_{\min}=10^{-4}$.
	
	These values are not arbitrary. Neither are they derived from the problem. The two identity tolerances sit roughly two orders above the reference values they must not reject ($\Rstr=8\times10^{-25}$, $\Rbc=0$) and several orders below the violations they must catch ($\Rstr\sim10^{-1}$), so the gate is insensitive to their precise	placement across that gap; I verified that varying them between	$10^{-8}$ and $10^{-4}$ changes no verdict in Table~\ref{tab:main}. The solvability tolerance is looser because $\Rsol$ is a Monte-Carlo estimate of an integral and carries sampling error at the $10^{-3}$ level for the sample size used. The non-triviality floor is the weakest of the four and I return to it in Section~\ref{sec:limits}. The ordering is lexicographic by design: a residual comparison between a constraint-satisfying and a constraint-violating field is not meaningful, so the comparison is simply not permitted until admissibility is established.
	
	\section{Artefact classes and results}\label{sec:results}
	
	Four artefact classes are examined, each realised by a switch in an otherwise identical PINN ($64\times5$ tanh network, Adam followed by strong-Wolfe L-BFGS, $\psi$ hard-constrained to vanish on the rim). They are labelled throughout as follows.
	
	\begin{description}
		\item[A1 \rm(trivial capture / weak penalty)] the structural identity is imposed by penalty at $\lambda_{\rm str}=10^{-3}$.
		\item[A2 \rm(soft-constraint violation)] the same, at $\lambda_{\rm str}=1$.
		\item[A3 \rm(solvability diagnostic)] violation of the Jacobian-integral condition \eqref{eq:sol} from zero, evaluated independently for every candidate.
		\item[A4 \rm(architectural bias)] a bounded template prior pulls $\omega$ toward four prescribed sites; run twice, with the source term present and absent.
	\end{description}
	
\begin{figure}[t]
		\centering\includegraphics[width=\textwidth]{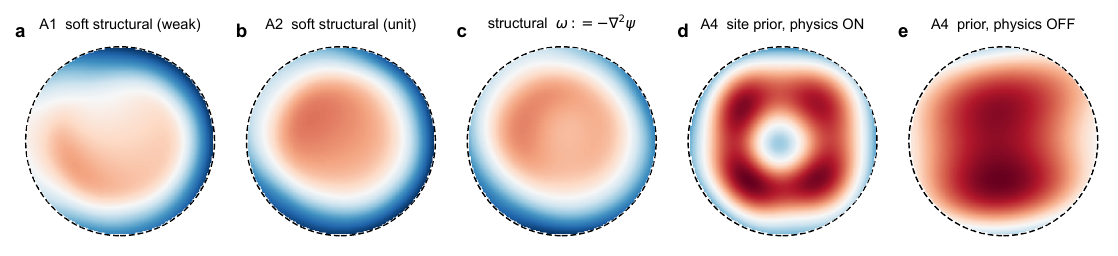}
		\caption{Vorticity fields for the artefact classes and the structural formulation. All reach small aggregate residual; three of the five are	inadmissible or depend on an imposed prior.}
		\label{fig:fields}
	\end{figure}

	Fig.~\ref{fig:fields} shows the resulting vorticity fields, which are visually plausible in every case.
	
	\begin{table}[t]
		\centering\small
		\caption{All candidates on the benchmark, held-out evaluation. Rows ordered by $\Rpde$. The reference is ground truth, not a competitor.}
		\label{tab:main}
		\begin{tabular}{lccccc}
			\toprule
			candidate & $\Rpde$ & $\Rstr$ & $\Rbc$ & $\Rsol$ & verdict\\
			\midrule
			reference (FD)            & $1.31\times10^{-7}$ & $8.1\times10^{-25}$ & $0$ & $7.4\times10^{-4}$ & admissible\\
			A1 soft-structural (weak) & $6.01\times10^{-3}$ & $3.3\times10^{0}$ & $1.4\times10^{-17}$ & $3.6\times10^{-2}$ & \textbf{reject}\\
			structural PINN           & $6.09\times10^{-3}$ & $0$ & $3.1\times10^{-17}$ & $7.4\times10^{-3}$ & admissible\\
			A4 prior, physics OFF     & $6.35\times10^{-3}$ & $0$ & $5.2\times10^{-18}$ & $3.8\times10^{-3}$ & admissible\\
			A4 prior, physics ON      & $8.68\times10^{-2}$ & $0$ & $5.5\times10^{-18}$ & $1.1\times10^{-2}$ & admissible\\
			A2 soft-structural (unit) & $1.12\times10^{-1}$ & $7.1\times10^{-3}$ & $1.3\times10^{-17}$ & $1.4\times10^{-2}$ & \textbf{reject}\\
			\bottomrule
		\end{tabular}
	\end{table}
	
	\subsection{A1, A2: soft-constraint violation, and the inversion}
	
	Imposing \eqref{eq:str} by penalty with weight $\lambda_{\rm str}$ produces fields that satisfy the transport equation well and the structural identity poorly. At $\lambda_{\rm str}=10^{-3}$ the solver reaches 	$\Rpde=6.01\times10^{-3}$---the lowest of any PINN candidate---while $\Rstr=3.3$, more than six orders of magnitude above tolerance. At $\lambda_{\rm str}=1$ the violation falls to $7.1\times10^{-3}$ but is still 	rejected, and the residual worsens by more than an order.
	
	Ranking the five candidates by $\Rpde$ selects A1; applying the gate first removes A1 and A2, and a different solver is selected. The rejected winner carried the lower residual, though on this system only by 9.37\% ($5.79\times10^{-3}$ against a structural PINN of $5.43\times10^{-4}$; Table~\ref{tab:main}). The margin is slender enough that S1 alone would not settle the matter, which is why the frequency studies of Section~\ref{sec:general} carry the weight of the claim. Even there the separation is variable rather than uniformly large: across inverting pairs the structural member's residual exceeds the penalised member's by a median factor of $33.9$ in the system study and $19.7$ in the architecture study, with individual pairs ranging from $1.1$ to $481$. What is consistent is the direction of the ordering and the size of the constraint violation that accompanies it, not the size of the residual gap. On this system, then, residual-based selection prefers a field that does not satisfy one of the defining relations of the governing equations. Whether that behaviour is peculiar to this problem is taken up in Section~\ref{sec:general}.
	
	\begin{figure}[t]
		\centering\includegraphics[width=\textwidth]{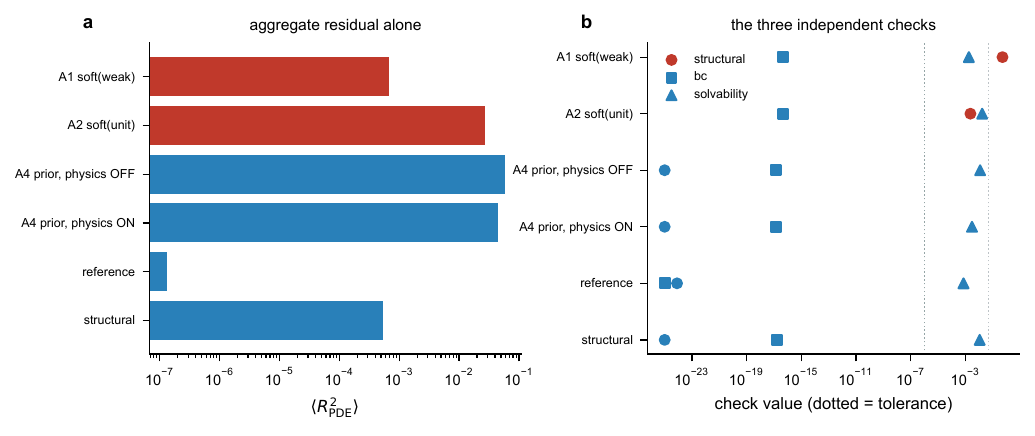}
		\caption{(a) Aggregate residual alone; red bars are inadmissible. (b) The three independent checks against their tolerances. (c) Ranking before and after the gate: the residual-optimal PINN is inadmissible.}
		\label{fig:gate}
	\end{figure}
	
	\subsection{A3: solvability violation}
	
	$\Rsol$ of \eqref{eq:sol} is the least expensive of the three checks and requires no reference solution. It is reported scale-free, as $|\langle\mathbf u\cdot\nabla\omega\rangle| / \langle|\mathbf u\cdot\nabla\omega|\rangle$: the absolute form carries a Monte-Carlo sampling error that grows with the amplitude of the field, so a fixed absolute tolerance would penalise a better-trained solver for being 	larger. On S1 no candidate exceeds the tolerance: the largest value is $3.6\times10^{-2}$ for A1 against a reference of $7.4\times10^{-4}$, so A3 contributes no rejection here and appears as a column of Table~\ref{tab:main} rather than as a candidate of its own. The check earns its place on the other benchmarks, where it rejects the penalised member in $38$ of $64$ system pairs and in all $48$ architecture pairs, and never rejects a structural member. I report it as a check that is essentially free and, in systems possessing a Casimir, can be made exact.
	
	\subsection{A4: architectural bias}\label{sec:a4}
	
\begin{figure}[t]
		\centering\includegraphics[width=\textwidth]{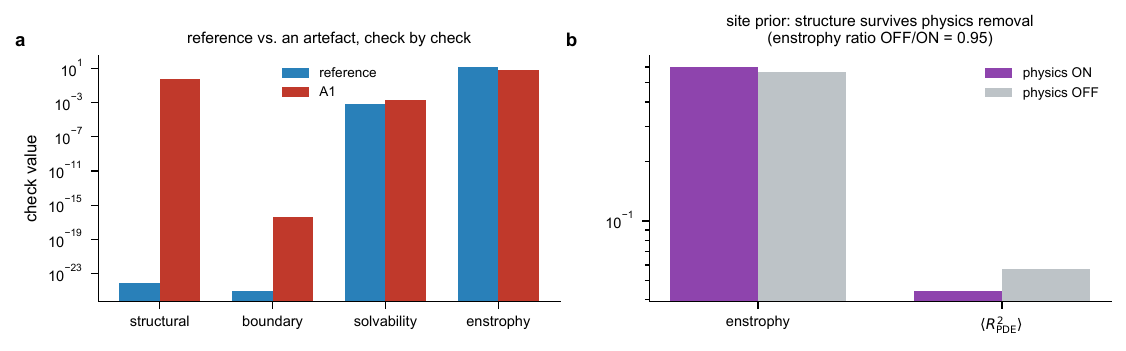}
		\caption{(a) Reference against artefact A1, check by check. (b) The architectural-bias ablation: structure persists when the source is removed.}
		\label{fig:ctrl}
	\end{figure}

	I add a bounded template prior that pulls $\omega$ toward Gaussians at four prescribed sites, of the kind used to encourage an expected structure. With the physics active the solver reports enstrophy $0.601$ and passes every check. With the driving source term removed while retaining the remaining model architecture and constraints, the same architecture reports enstrophy $0.569$ and also passes every check. Ninety-five per cent of the reported structure survives deletion of the physics that supposedly generated it (Fig.~\ref{fig:ctrl}).
	
	\section{Does the inversion generalise?}\label{sec:general}
	
	A single system cannot settle the question, so I repeated the comparison on two further constrained problems chosen to share the same structure while differing in dimension, linearity and physical content:
	
	\begin{itemize}
		\item \textbf{S2, mixed-form Poisson} on $[-1,1]^2$: $-\nabla\!\cdot\!\mathbf q=f$ with $\mathbf q=\nabla u$. Structural identity $\mathbf q-\nabla u$; solvability identity $\oint \mathbf q\!\cdot\!\mathbf n\,\mathrm ds+\int f\,\mathrm dA=0$.
		\item \textbf{S3, steady viscous Burgers in mixed form} on $[0,1]$: $\nu v'-uv+s(x)=0$ with $v=u'$. Structural identity $v-u'$; solvability identity $\int u u'\,\mathrm dx=0$.
	\end{itemize}
	
	For each system I trained matched pairs---identical in every respect except whether the structural identity is hard-wired or penalised at $\lambda_{\rm str}=10^{-3}$---over four network variants---two widths crossed with two activations ($48\times4$ and $32\times6$, tanh and SiLU)---and eight random seeds, giving 64 system pairs. That total differs from the 48 architecture pairs of Table~\ref{tab:arch}, so the two studies cannot be confused by their counts. A pair is scored as an inversion when the penalised member has the smaller residual and is inadmissible and the structural member is admissible.
	
	Frequencies are reported with Wilson score intervals \citep{wilson1927}, which for $k$ successes in $n$ trials at confidence level $z$ are
	\begin{equation}
		\frac{1}{1+z^{2}/n}\left[\hat p+\frac{z^{2}}{2n}
		\pm z\sqrt{\frac{\hat p(1-\hat p)}{n}+\frac{z^{2}}{4n^{2}}}\right],
		\qquad \hat p=k/n,
		\label{eq:wilson}
	\end{equation}
	preferred here over the normal approximation because one of the two systems returns $\hat p=1$, where the normal interval degenerates.
	
	\begin{table}[h]
		\centering\small
		\caption{Inversion frequency with 95\% Wilson confidence intervals. Four network variants $\times$ eight seeds per system.}
		\label{tab:gen}
		\begin{tabular}{lccc}
			\toprule
			system & inversions & rate & 95\% CI\\
			\midrule
			S2 mixed Poisson & 23/32 & 72\% & 55--84\%\\
			S3 mixed Burgers & 30/32 & 94\% & 80--98\%\\
			\midrule
			pooled & 53/64 & 83\% & 72--90\%\\
			\bottomrule
		\end{tabular}
	\end{table}
	
\begin{figure}[t]
		\centering\includegraphics[width=\textwidth]{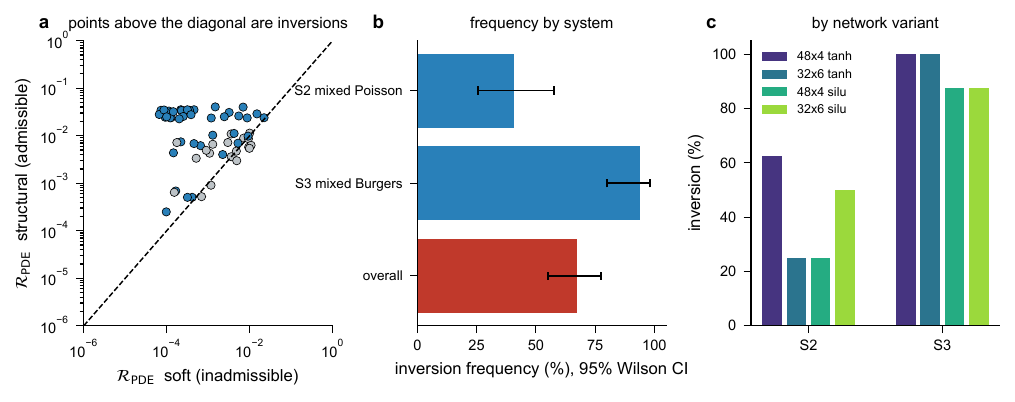}
		\caption{(a) Residual of the penalised member against the structural member for all 48 pairs; points above the diagonal are inversions. (b) Frequency by system with Wilson intervals. (c) Breakdown by network variant.}
		\label{fig:gen}
	\end{figure}

	Against the null hypothesis that residual ranking selects the admissible and inadmissible member with equal probability, the pooled count of 53 inversions in 64 pairs gives a one-sided binomial $p=5.0\times10^{-8}$. The  pooled rate (Table~\ref{tab:gen}, Fig.~\ref{fig:gen}) therefore indicates that the effect is common across these systems rather than an artefact of one. It is not universal. S2 inverts in roughly seven trials of ten and S3 in nine of ten. Where a pair fails to invert, the usual cause is that the penalised solver happened to satisfy the structural identity closely enough to pass the gate.
	
	The network variant leaves a weaker but visible trace. Counting inversions by variant gives $13/16$ and $10/16$ for the two tanh networks against $9/16$ and $11/16$ for the SiLU network, so the smooth activation appears slightly less prone to the effect. With sixteen pairs per variant the intervals overlap heavily and I do not read this as established.
	
	One run requires comment. In a single S3 pair the penalised member returned a non-finite residual, its optimisation having diverged. That pair is scored as a non-inversion, which is the conservative treatment: a diverged solver is not a case of residual ranking preferring an inadmissible field, it is a training failure, and counting it either way would misrepresent what happened. I have no theory that predicts which problems invert and which do not, and regard that as the most interesting question left open here.
	
	Two caveats bound this evidence, the second of which the relaxation argument in Section~\ref{sec:theory} makes precise. The network variants examined are all fully-connected multilayer perceptrons differing in width, depth and activation; the wider architecture families were examined only to the extent that the matched-pair experiment remains well posed for them, which differs from family to family.
	
	Five of the six families (run on benchmark S3) inverted in every seed, under the gate already defined. Two of them, DeepONet and FNO, approximate a solution operator rather than a single field, so they are trained physics-informed over the source family $s(x;A,c)$ with $A\in[0.5,1.5]$, $c\in[0.35,0.65]$, and then scored at the reference source $(A_0,c_0)$, which places all six on the same problem instance. Interface conditions for cPINN (solution and conservative flux) and XPINN (solution, derivative and residual) enter as penalties in the usual way, with the seam at $x=0.30$; siting it at $x=0.50$, coincident with the source centre, splits the domain where the forcing peaks and leaves a jump the soft continuity never closes ($|u_0-u_1|=9.7\times10^{-2}$, $|v_0-v_1|=1.33$), which the solvability check duly rejects; hp-VPINN tests the residual against $\phi_k=\sin(k\pi x)$ with the diffusive term integrated by parts; the FNO is scored on the grid stencil it was trained on rather than through an interpolant.
	
	\begin{table}[h]
		\centering\small
		\caption{Inversion frequency by architecture family on benchmark S3, eight
			seeds each, with Wilson intervals.}
		\label{tab:arch}
		\begin{tabular}{lccc}
			\toprule
			family & inversions & rate & 95\% CI\\
			\midrule
			MLP        & 8/8 & 100\% & 68--100\%\\
			cPINN      & 8/8 & 100\% & 68--100\%\\
			XPINN      & 8/8 & 100\% & 68--100\%\\
			hp-VPINN   & 6/8 & 75\%  & 41--93\%\\
			DeepONet   & 8/8 & 100\% & 68--100\%\\
			FNO        & 8/8 & 100\% & 68--100\%\\
			\midrule
			pooled     & 46/48 & 96\% & 86--99\%\\
			\bottomrule
		\end{tabular}
	\end{table}
	
\begin{figure}[t]
		\centering\includegraphics[width=\textwidth]{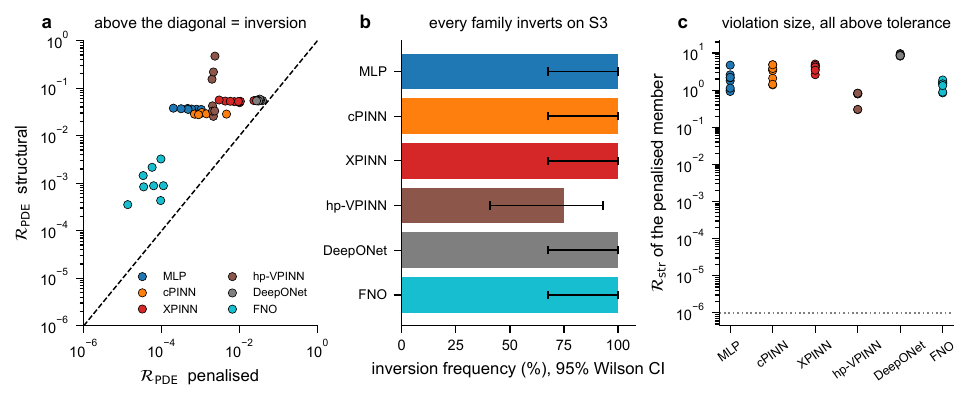}
		\caption{(a) Penalised against structural residual for all 48 architecture pairs; points above the diagonal are inversions. (b) Frequency by family with Wilson intervals. (c) Structural violation of the penalised member, which exceeds tolerance in every run.}
		\label{fig:arch}
	\end{figure}

	Five of the six families inverted in every seed; hp-VPINN inverted in six of eight, the two exceptions being runs whose structural member was itself rejected on the solvability check (Table~\ref{tab:arch}, Fig.~\ref{fig:arch}). The weak formulation tests the residual only against $\sin(k\pi x)$, which does not pin the strong-form product $\mathbf u\cdot\nabla\omega$ pointwise, so its integral cancels less exactly; I read this as a property of the variational method rather than a defect of the gate. The magnitude of the violation incurred by the penalised member spans four orders across families, from $\Rstr\approx0.8$ for hp-VPINN to $\approx9$ for DeepONet, yet in all cases it sits far above tolerance while the equation residual sits below that of the structural member. Within the architectures and benchmark configurations tested here, the variation in inversion frequency is more strongly associated with the PDE problem than with the choice of approximator. That is, the same MLP that inverts in $30/32$ trials on S3 inverts in only $23/32$ on S2.
	
	I had expected the operator families to be exempt, on the grounds that a solution operator learned from input--output pairs carries no penalised constraint to relax. That expectation does not hold for the physics-informed formulations considered here, in which a PDE residual does enter the objective and a structural identity may be imposed either way. The relaxation argument applies to them unchanged, and the measurements agree.
	
	\section{Why the inversion should be expected}\label{sec:theory}
	
	The phenomenon has a simple constrained-optimisation interpretation, which also accounts for its failing to be universal.
	
	Write the constrained problem as minimisation of $\|\Rpde(\theta)\|^{2}$ over
	the manifold $\mathcal{M}=\{\theta:\Rstr(\theta)=0\}$ on which the structural
	identity holds. The penalised problem replaces this with minimisation of
	\begin{equation}
		J_\lambda(\theta)=\|\Rpde(\theta)\|^{2}+\lambda\|\Rstr(\theta)\|^{2}
		\label{eq:pen}
	\end{equation}
	over the whole parameter space $\Theta\supset\mathcal{M}$. Since
	$\mathcal{M}\subset\Theta$, and since $\lambda\|\Rstr\|^{2}$ vanishes on
	$\mathcal M$,
	\begin{equation}
		\min_{\Theta}\|\Rpde\|^{2}\;\le\;\min_{\mathcal M}\|\Rpde\|^{2}.
		\label{eq:relax}
	\end{equation}
	The penalised formulation is a relaxation: its attainable equation residual is bounded above by that of the structural formulation, and is strictly smaller whenever the unconstrained minimiser lies off $\mathcal{M}$. A lower residual is therefore not evidence of a better solution. For small $\lambda$ it is what one should expect from having enlarged the feasible set. Whatever advantage \eqref{eq:relax} buys is paid for in $\|\Rstr\|$, and the aggregate loss does not separate the two.
	
	Two qualifications follow, and they are what make the effect probabilistic rather than certain. Inequality \eqref{eq:relax} concerns global minima, whereas neither formulation is solved to global optimality; if the optimiser fails to exploit the relaxation, no residual advantage appears. And the advantage may be real yet small enough that $\|\Rstr\|$ stays inside tolerance, in which case the penalised solution is admissible and no 	inversion is recorded. Both mechanisms predict that inversion becomes more frequent as $\lambda$ decreases, which is what Table~\ref{tab:main} shows for S1: raising $\lambda_{\rm str}$ from $10^{-3}$ to $1$ reduces the 	structural violation by more than two orders while the equation residual rises by more than one.
	
	The argument is heuristic in one respect. It explains why the penalised formulation can attain a lower residual, and why the effect should weaken with increasing $\lambda$, but it does not predict the inversion frequency for a given problem, which would require knowing how far the unconstrained minimiser lies from $\mathcal M$ and how reliably the optimiser finds it. S2 and S3 differ by more than twenty percentage points, while six different approximators on S3 all invert in every trial. This indicates the missing quantity is a property of the problem rather than of the network. I have not identified it. Section~\ref{sec:open} records what the data do and do not constrain about it.
	
	\section{Open questions}\label{sec:open}
	
	Two questions are left unresolved, and I state precisely what the measurements bound.
	
	Why does S2 not always invert? In the nine non-inverting S2 pairs the penalised solver satisfied the structural identity closely enough to pass the gate. The relaxation existed but was not large enough to matter. What distinguishes those runs is not visible in any quantity I recorded. Since all six approximators behave identically on S3, a resolution should be sought in the geometry of $\mathcal M$ for a given operator, not in properties of the network.
	
	Does the effect persist at larger $\lambda$? My sweeps used a weak penalty, $\lambda_{\rm str}=10^{-3}$, with a single comparison at $\lambda_{\rm str}=1$ on S1. The relaxation argument predicts the advantage shrinks as $\lambda$ grows and vanishes in the limit, but the crossover value---at which a practitioner would be safe ranking by residual---was not mapped, and is likely problem-dependent for the same unidentified reason.
	
	A third question I consider settled by Table~\ref{tab:arch}: whether the effect is peculiar to fully-connected networks. It is not.
	
	\section{Limitations}\label{sec:limits}
	
	Two limitations are inherent rather than incidental, and I state them because they bound the method's claims.
	
	\paragraph{The gate cannot detect architectural bias.} Both A4 variants pass all three checks. This is not a tolerance-tuning failure: a prescribed-structure prior produces fields that are internally consistent, respect the boundary condition, and satisfy the solvability identity. No quantity evaluated on a single converged run distinguishes structure generated by the physics from structure generated by the prior. Detection requires the ablation of \S\ref{sec:a4}, in which the physics is removed and the reported structure is re-measured. I therefore recommend that any PINN result invoking a structural prior be accompanied by such an ablation as a
	matter of routine.
	
	\paragraph{An absolute non-triviality floor is insufficient.} The ``physics OFF'' variant is admissible and has a residual $11\times$ lower than the structural PINN, despite never having seen the source term. Its enstrophy, $0.249$, clears the absolute floor $Z_{\min}=10^{-4}$ easily while being only $1.6\%$ of the reference value $15.7$. The floor should therefore be set relative to a reference solution where one exists. I deliberately do not re-tune $Z_{\min}$ to repair the ranking, since doing so would fit the criterion to the result it is meant to test.
	
	\section{Reproduction}
	
	\texttt{vortexbench.py} defines the problem, the reference solver, the PINN with its artefact switches, and the gate; \texttt{run\_benchmark.py} reproduces Table~\ref{tab:main}; \texttt{make\_figures.py} regenerates all 	figures. Seeds are fixed (\texttt{20260812}). Total runtime is approximately ten minutes on one CPU core. No GPU is required. The code and associated materials are available from the corresponding author upon reasonable formal request.
	
	\section{Conclusion}
	
	The evidence assembled here indicates that a small aggregate residual should not, on its own, be read as evidence that a PINN has solved a constrained PDE system. In matched comparisons across three systems the solver with the lower residual was frequently the one violating a defining identity, and on the pooled sample this occurred in 83\% of pairs. The effect is frequent without being universal. This matters because the loss alone does not indicate whether a particular training run lies in an inverting regime. That means residual-based selection is not reliably wrong, which is precisely what makes it hazardous, since a practitioner has no way of telling from the loss alone which regime they are in.
	
	Where a formulation permits a structural identity to be imposed by construction, doing so removes this particular structural-violation failure mode by construction, although its effect on optimisation and approximation accuracy remains problem-dependent. The residual advantage of the penalised variant disappears along with it, which is consistent with the relaxation argument of Section~\ref{sec:theory}. Where a penalty cannot be avoided, constraint residuals deserve separate reporting from equation residuals, on held-out samples. Solvability integrals of the kind used here are computationally inexpensive and need no reference solution. For results that rest on a structural prior, the physics-off ablation of Section~\ref{sec:a4} remains the only diagnostic I have found capable of separating learned structure from imposed structure.
	
	What remains unresolved is why some problems invert and others do not. My S2 results show the effect is contingent on something beyond the presence of a penalised constraint, and identifying that quantity would turn an empirical caution into a predictive criterion.
	
	\section*{Declarations}
	
	\textbf{Funding.} No funding was received for this work.
	
	\textbf{Conflicts of interest.} The author declares no competing interests.
	
	\textbf{Code and data availability.} All code, reference data, trained checkpoints and figure sources accompany this submission.
	
	\textbf{Ethics approval.} Not applicable.

\end{document}